\documentclass[letter]{aa} 

\usepackage{graphicx}
\usepackage{txfonts}
\usepackage[colorlinks=true,linkcolor=red,citecolor=blue]{hyperref}

\DeclareUnicodeCharacter{2212}{-}

\defcitealias{gilli_2019}{G19}

\begin{document}

   \title{Discovery of molecular gas fueling galaxy growth in a protocluster at z=1.7}


   \author{Q. D'Amato \inst{1,2}\thanks{\email{quirino.damato2@unibo.it}}
          \and
          R. Gilli\inst{3}
          \and
          I. Prandoni\inst{1}
          \and
          C. Vignali\inst{2,3}
          \and
          M. Massardi\inst{4}
          \and
          M. Mignoli\inst{3}
          \and
          O. Cucciati\inst{3}
          \and
          T. Morishita\inst{5}
          \and
          R. Decarli\inst{3}
          \and
          M. Brusa\inst{2,3}
          \and
          F. Calura\inst{3}
          \and
          B. Balmaverde\inst{6}
          \and
          M. Chiaberge\inst{5}
          \and
          E. Liuzzo\inst{4}
          \and
          R. Nanni \inst{7,3}
          \and
          A. Peca\inst{8,3}
          \and
          A. Pensabene\inst{3,2}
          \and
          P. Tozzi\inst{9}
          \and
          C. Norman\inst{5}
          }

   \institute{INAF/IRA, Istituto di Radioastronomia, Via Piero Gobetti 101, 40129, Bologna, Italy
         \and
             Dipartimento di Fisica e Astronomia dell’Università degli Studi di Bologna, via P. Gobetti 93/2, 40129 Bologna, Italy
         \and  
             INAF/OAS, Osservatorio di Astrofisica e Scienza dello Spazio di Bologna, via P. Gobetti 93/3, 40129 Bologna, Italy
         \and  
             INAF, Istituto di Radioastronomia - Italian ALMA Regional Center (ARC), Via Piero Gobetti 101, 40129, Bologna, Italy
         \and
             Space Telescope Science Institute, 3700 San Martin Drive, Baltimore, MD 21218, USA
         \and
             INAF, Osservatorio Astronomico di Torino, Via Osservatorio 20, 10025, Pino Torinese (TO), Italy
         \and
             Department of Physics, University of California, Santa Barbara, CA 93106-9530, USA
         \and
             Department of physics, University of Miami, Coral Gables, FL 33124, USA
         \and
             INAF, Osservatorio Astrofisico di Arcetri, Largo E. Fermi 5, I-50125 Firenze, Italy
         }

\date{Received XXX; accepted XXX}

 
  \abstract
{Based on ALMA Band 3 observations of the CO(2$\rightarrow$1) line transition, we report the discovery of three new gas-rich ($M_{H2} \sim 1.5-4.8 \times 10^{10}~M_\odot$) galaxies in an overdense region at z=1.7 that already contains eight spectroscopically confirmed members. This leads to a total of 11 confirmed overdensity members within a projected distance of $\sim$ 1.15 Mpc and in a redshift range of $\Delta$z = 0.012. Under simple assumptions, we estimate that the system has a total mass of $\geq 3-6 \times 10^{13}~M_\odot$, and show that it will likely evolve into a $\gtrsim 10^{14}~M_\odot$ cluster at z = 0. The overdensity includes a powerful Compton-thick Fanaroff-Riley type II (FRII) radio galaxy, around which we discovered a large molecular gas reservoir ($M_{H2} \sim 2 \times 10^{11}~M_\odot$). We fit the FRII resolved CO emission with a 2D Gaussian model with a major (minor) axis of $\sim$ 27 ($\sim$17) kpc, which is a factor of $\sim$3 larger than the optical rest-frame emission. Under the assumption of a simple edge-on disk morphology, we find that the galaxy interstellar medium produces a column density toward the nucleus of $\sim 5.5 \times 10^{23}~ \mathrm{cm^{-2}}$. A dense interstellar medium like this may then contribute significantly to the total nuclear obscuration measured in the X-rays ($N_{H,X} \sim 1.5 \times 10^{24} ~\mathrm{cm^{-2}}$) in addition to a small, paresec-scale absorber around the central engine. The velocity map of this source unveils a rotational motion of the gas that is perpendicular to the radio jets.\\
All ALMA sources have a dust-reddened counterpart in deep Hubble Space Telescope images (bands $i$, $z$, $H$), while we do not detect any molecular gas reservoir around the known UV-bright, star-forming members discovered by MUSE. This highlights the capability of ALMA of tracing gas-rich members of the overdensity. For the MUSE sources, we derive 3$\sigma$ upper limits to the molecular gas mass of $M_{H_2}\leq 2.8-4.8 \times 10^{10}~\mathrm{M_\odot}$ . We derive star formation rates in the range ${\sim}5-100~M_{\odot}$/yr for the three new ALMA sources.\\
The FRII is located at the center of the projected spatial distribution of the structure members, and its velocity offset from the peak of the redshift distribution is well within the velocity dispersion of the structure. All this, coupled with the large amount of gas around the FRII, its stellar mass of ${\sim}3 \times 10^{11} M_\odot$, star formation rate of $\sim 200-600~M_{\odot}$/yr, and powerful radio-to-X-ray emission, suggests that this source is the likely progenitor of the future brightest cluster galaxy.
}
   
   \maketitle
%

\section{Introduction}
Galaxy clusters are the largest virialized structures in the Universe. The physical processes leading to their formation are mostly investigated through numerical simulations, which can now be tested through direct observations of cluster progenitors. Cosmological simulations show that most of the large-scale structure assembly takes place at a redshift between z${\sim}$4 and z${\sim}$1 \citep{boylan_2009}. At this epoch, the cosmic star formation rate (SFR) and black hole accretion peak, and most of the galaxy stellar mass is built \citep{madau_2014}.
Protoclusters (i.e., large-scale nonvirialized structures that will collapse into galaxy clusters of at least $10^{14}~\mathrm{M_\odot}$; \citeauthor{bower_2004} \citeyear{bower_2004}) represent the early stages of this assembly (see \citeauthor{overzier_2016} \citeyear{overzier_2016} for a review). While their gravitational collapse can be studied theoretically in numerical cosmological simulations \cite[e.g.,][]{pillepitch_2018}, many details of the ongoing processes regulating their evolution remain mostly unknown, especially prior to virialization, because only a few systems are known \citep{chiang_2013}.\\
 \citeauthor{gilli_2019} (\citeyear{gilli_2019}, hereafter \citetalias{gilli_2019}) reported the discovery of a galaxy overdensity around a powerful (X-ray luminosity: $L_{2-10~ \mathrm{keV}}{\sim} 1.3\times 10^{44}$ erg/s; total rest-frame extended radio power: $P_{408~\mathrm{MHz}} {\sim}$  $10^{26}$  $\mathrm{W~Hz^{-1}~sr^{-1}}$), massive (stellar mass $M_{\ast}$ ${\sim}$ $3 \times 10^{11} M_\odot$), Compton-thick (X-ray derived column density $N_{H,X}{\sim}10^{24}~\mathrm{cm^{-2}}$) Fanaroff-Riley type II \cite[FRII;][]{fanaroff_1974} high-redshift radio galaxy (HzRG) at z=1.6987, located at the center of the so-called  J1030+0524 field\footnote{This $\sim$30'$\times$30' field, centered around the z=6.3 Sloan Digital Sky Survey J1030+0524 quasar, is targeted by deep and wide surveys that span in frequency from the X-ray to the radio band, as a result of major observational programs led by our group (\url{http://j1030-field.oas.inaf.it}).}. The FRII radio galaxy was first revealed by \cite{petric_2003} from deep 1.4 GHz Very Large Array (VLA) observations of the field. \cite{nanni_2018} reprocessed the data and published the first analysis (extension and morphology) of the FRII radio galaxy (shown as white contours in the left panel of Fig. \ref{fig:HST+srcs}). By means of deep (480 ks) \textit{Chandra} observations of the field, they also discovered an X-ray diffuse emission, mostly coinciding with the eastern lobe and central nucleus of the FRII (orange contours in Fig. \ref{fig:HST+srcs}).
Based on spectroscopic data obtained with the Very Large Telescope Multi Unit Spectroscopic Explorer (MUSE) and Large Binocular Telescope Utility Camera in the Infrared (LUCI) instruments, \citetalias{gilli_2019} identified eight star-forming galaxies (SFGs) distributed within a projected distance of 800 kpc from the FRII host galaxy in the redshift range 1.6871 -- 1.6987 (including the FRII core, indicated as \object{$l1$} in the left panel of Fig. \ref{fig:HST+srcs}, yellow circle). MUSE galaxies show blue spectra (Fig. 5 of \citetalias{gilli_2019}), whose UV emission is dominated by star formation activity (SFR $\sim 8-60~\mathrm{M_{\odot}}$/yr; see Table 1 and Table 3 of \citetalias{gilli_2019} for more details). Remarkably, four out of eight of the discovered members (indicated as \object{$m1$} - \object{$m4$} in the left panel of Fig. \ref{fig:HST+srcs}, green circles) lie in an arc-like shape at the edge of the X-ray diffuse emission around the eastern lobe of the FRII. \citetalias{gilli_2019} proposed that this diffuse emission originates from an expanding bubble of gas that is shock-heated by the FRII jet, triggering the SF observed in \object{$m1$} - \object{$m4$}. If confirmed, this would be the first evidence of positive feedback by active galactic nuclei (AGN) on multiple galaxies at hundreds of kiloparsec distance.
In order to detect other overdensity members and possibly unveil large gas reservoirs, we obtained new Atacama Large Millimeter/submillimeter Array (ALMA) observations of the region (${\sim}4~ \mathrm{arcmin}^2$) around the radio galaxy, aiming to detect the carbon monoxide CO(2$\rightarrow$1) transition line.
Here we report the discovery of three new gas-rich members of the overdensity ($M_{H2} \sim 1.5-4.8 \times 10^{10}~M_\odot$) and of a large molecular gas reservoir (${\sim} 2 \times10^{11}~\mathrm{M_\odot}$) around the FRII core.\\
Throughout the work we adopt a concordance $\Lambda$CDM cosmology with $\mathrm{H_0} = 70~\mathrm{km~s^{-1}~Mpc^{-1}}$, $\Omega_{\mathrm{M}} = 0.3$, and $\Omega_{\mathrm{\Lambda}} = 0.7$, in agreement with the \textit{Planck 2015} results \citep{PLANCK_2016}.
The angular scale and luminosity distance at z=1.7 are 8.46 kpc/arcsec and 12.7 Gpc, respectively.

\section{Observations and data analysis}
\label{sec:obs}
We performed a three-pointing mosaic observations with ALMA Band 3 (84 -- 116 GHz) during Cycle 6 (project ID: {2018.}1.{01601}.S). The field of view (FoV) of the observation is shown by the solid magenta line in Fig. \ref{fig:HST+srcs}. This corresponds to the half primary beam width (\textit{HPBW}) at the tuning frequency of the observations.
The $2 \times 4$ GHz spectral windows (\textit{spws}) cover the 84.1 -- 87.7 GHz and 98.0 -- 99.9 GHz spectral ranges. Each \textit{spw} is sampled in 1920 channels of 976.5 kHz width, corresponding to 3.2 km/s at the mean frequency of the \textit{spws}. The observations consist of six ${\sim}50.5$ min execution blocks carried out on $18^{\mathrm{}}$ (two execution blocks), $28^{\mathrm{}}$ (one execution block), and $31^{\mathrm{}}$ (three execution blocks) December 2018. The quasars J1058+0133 and J1038+0512 served as flux and bandpass and phase calibrator, respectively.
We performed the data calibration and flagging through the calibration pipeline of the Common Astronomy Software Applications (CASA) package \cite[version 5.4.0-70;][]{mcmullin_2007}.

\subsection{Imaging}

\citetalias{gilli_2019} showed that the protocluster members can be found at large projected separation (${\sim}800$ kpc for \object{$l2$}; outside the region shown in Fig. \ref{fig:HST+srcs}) and large velocity offset (${\sim}1300$ km/s for \object{$m6$}) from the FRII radio galaxy. In order to detect new members of the overdensity, we therefore produced a datacube (with the {\sc{tclean}} CASA task) of the entire FoV covered by the mosaic, within ${\pm 3000}$ km/s from the FRII optical rest-frequency (85.426 GHz). The continuum, estimated from the remaining observed spectral range, was fit using a polynomial and subtracted from the data set through the {\sc{uvcontsub}} CASA task.
A 30 km/s channel width was chosen for the datacube: this allowed us to resolve the relevant kinematic features of the sources, with a high enough signal-to-noise ratio (S/N) per channel to properly constrain the spectral parameters.
The robust weighting was set to 0.5, corresponding to a restoring beam with a major (minor) axis of 2.161 (1.655) arcsec.\\

\subsection{S/N cube and detections}
\label{subsec:SNR_cube}
\begin{figure*}[h!]
        \centering
\includegraphics[scale=0.6]{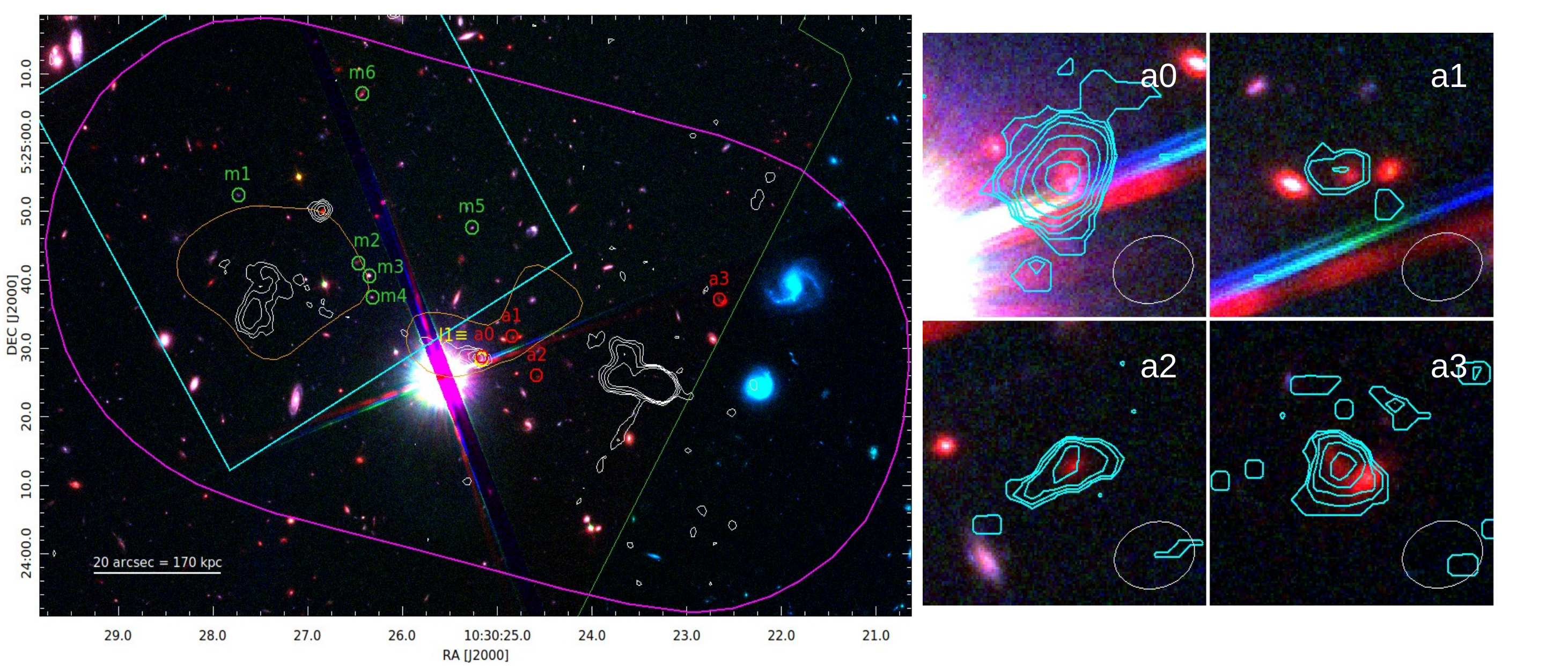}

        \caption{\textit{Left:} Composed r-g-b image of the field around the FRII radio galaxy. Red, green, and blue channels are the HST F160w, F850lp, and F775w filters. Green and red circles mark the MUSE (\object{$m1$}-\object{$m6$}) and ALMA (\object{$a0$}-\object{$a3$}) sources, respectively. \object{$a0$} (the FRII core) corresponds to the LUCI source \object{$l1$}, marked in yellow. A second LUCI source, \object{$l2$}, is located ${\sim}$1.5 arcmin southeast of \object{$a0$}, outside the image. VLA contours at 1.4 GHz are overlaid in white, starting at the ${\sim}3 \sigma$ level emission and increasing with a $\sqrt{3}$ geometric progression. The orange contours mark the ${\sim}2.5 \sigma$ level of the diffuse X-ray emission (0.5-7 keV). The solid green polygon delimits the region covered simultaneously by all HST filters. The solid magenta line and cyan box are the ALMA \textit{HPBW} and MUSE FoV, respectively. The solid white line in the bottom left corner is the angular and physical scale. \textit{Right}: ${\sim}7 \times 7$ $\mathrm{arcsec^2}$ (${\sim}60 \times 60$ $\mathrm{kpc^2}$) cutouts of the composed HST r-g-b image, centered on the ALMA detected sources (named as labeled). The cyan contours represent the 3$\sigma$ moments 0 map of our observations (see Sect. \ref{subsec:spec_an_mom}), starting at the ${\sim}3 \sigma$ level emission and increasing with a $\sqrt{3}$ geometric progression. The white ellipse in the bottom right corners represents the restoring beam.
        }
        
        \label{fig:HST+srcs}
\end{figure*}

We performed the source detections, moments, and spectral analysis exploiting the S/N cube, in which each pixel has the value of the S/N of the pixel in the original datacube, calculated as described in Appendix \ref{app:det_code}. We developed a code that aims to detect emission lines in the S/N cube without prior knowledge of their spectral (and spatial) position and properties on the basis of spectral, spatial and reliability criteria. In this work we report only the four secure detections found by the code, which all show clear strong CO(2$\rightarrow$1) emission (Fig. \ref{fig:spectra}; see Appendix \ref{app:det_code} for details about the detections). As for \object{$a1$}, \object{$a2$}, and \object{$a3$}, based on the CO luminosity functions derived by the ASPECS project \citep{decarli_2019} and the sensitivity of our observations, we estimate that the number of interlopers expected in the observed volume is only 0.09 sources, a factor 2.8 lower than the expected CO(2$\rightarrow$1) emitters; in any case, we find that the probability of observing three CO(2$\rightarrow$1) transitions in a blank field would be very low (P = $2\times 10^{-3}$, assuming a Poisson distribution). Thus, the presence of a known overdensity at z=1.7 in the field naturally accounts for the observed number of lines as CO(2$\rightarrow$1) transitions.
The positions of the ALMA detections are reported in Table \ref{tab:sample} and indicated in the composed r-g-b Hubble Space Telescope image (HST; left panel of Fig. \ref{fig:HST+srcs}) by the red circles. All galaxies have an optical counterpart; their HST $H$-band magnitude $H_{AB}$ and color index $z-H$ are reported in Table \ref{tab:sample}. The description and reduction of the HST data will be reported elsewhere (Morishita et al. in prep.)

\begin{table}
\caption{\label{tab:sample} Summary of the ALMA detections.
}
\centering
\resizebox{\hsize}{!}{
\begin{tabular}{ccccc}
\hline\hline

ID                  & RA(J2000)                 & DEC(J2000)                                     & $H_{AB}$    &    $z-H$       \\
(1)                 & (2)                & (3)                                     &   (4)       &     (5)        \\
\hline
\object{$a0$} & $10^h 30^m 25^s.16 \pm 0^s.13$ & $+5^{\circ} 24' 28''.69 \pm 0^s.13$   &    22.43      &     2.3        \\
\object{$a1$} & $10^h 30^m 24^s.84 \pm 0^s.23$ & $+5^{\circ} 24' 31''.68 \pm 0^s.23$   &    23.25      &     2.2        \\
\object{$a2$} & $10^h 30^m 24^s.58 \pm 0^s.31$ & $+5^{\circ} 24' 25''.97 \pm 0^s.31$   &    23.54      &     3.1        \\
\object{$a3$} & $10^h 30^m 22^s.65 \pm 0^s.23$ & $+5^{\circ} 24' 37''.08 \pm 0^s.23$   &    23.20 (21.7)      &     2.3 (3.2)        \\

\hline
\end{tabular}
}
\tablefoot{
(1) Source ID.
(2) Source right ascension (RA) and (3) declination (DEC).
(4) HST $H$-band (typical 1$\sigma$ uncertainty: 0.02) and (5) $z-H$ color index (typical 1$\sigma$ uncertainty: 0.25). \\
The 1$\sigma$ uncertainty on the detection position is set to $0.5$ $\times$ $\langle FWHM \rangle$ $\times$ $\mathrm{S/N}^{-1}$ \citep{papadopoulos_2008}, where $\langle FWHM \rangle$ is the mean of the major and minor \textit{FWHM} of the beam and the S/N is that of the peak pixel.
We note that \object{$a3$} appears to be a blend of two galaxies at the higher spatial resolution of the HST images (Fig. \ref{fig:HST+srcs}, right); we report the magnitude and color index of the source corresponding to the bulk of CO emission (that to the east) and in parentheses the values of the other source (that to the west).
}
\end{table}

\subsection{Spectral analysis and moments}
\label{subsec:spec_an_mom}

\begin{figure*}[h!]
        \centering
\includegraphics[scale=0.3]{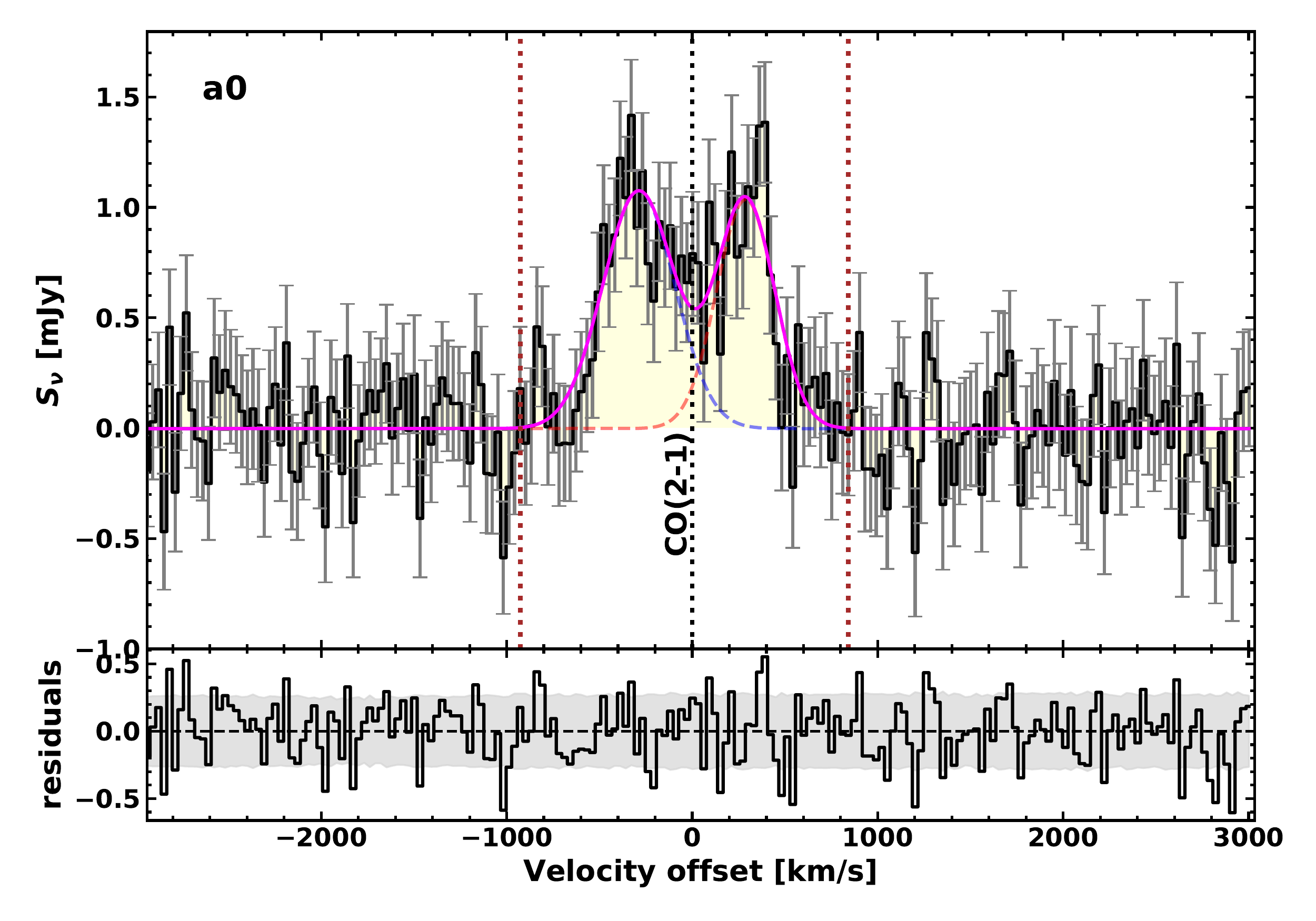}
\includegraphics[scale=0.3]{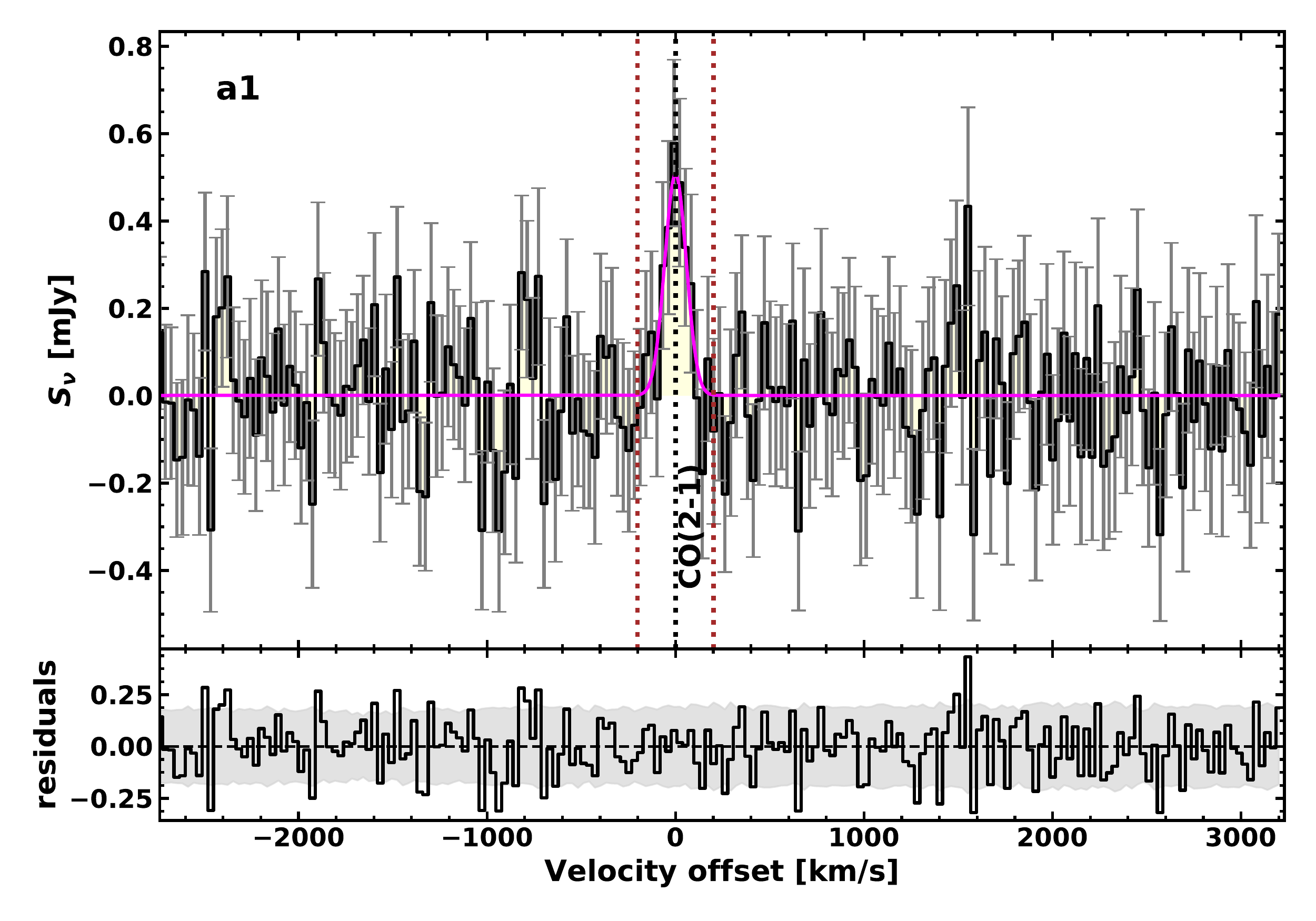}
\includegraphics[scale=0.3]{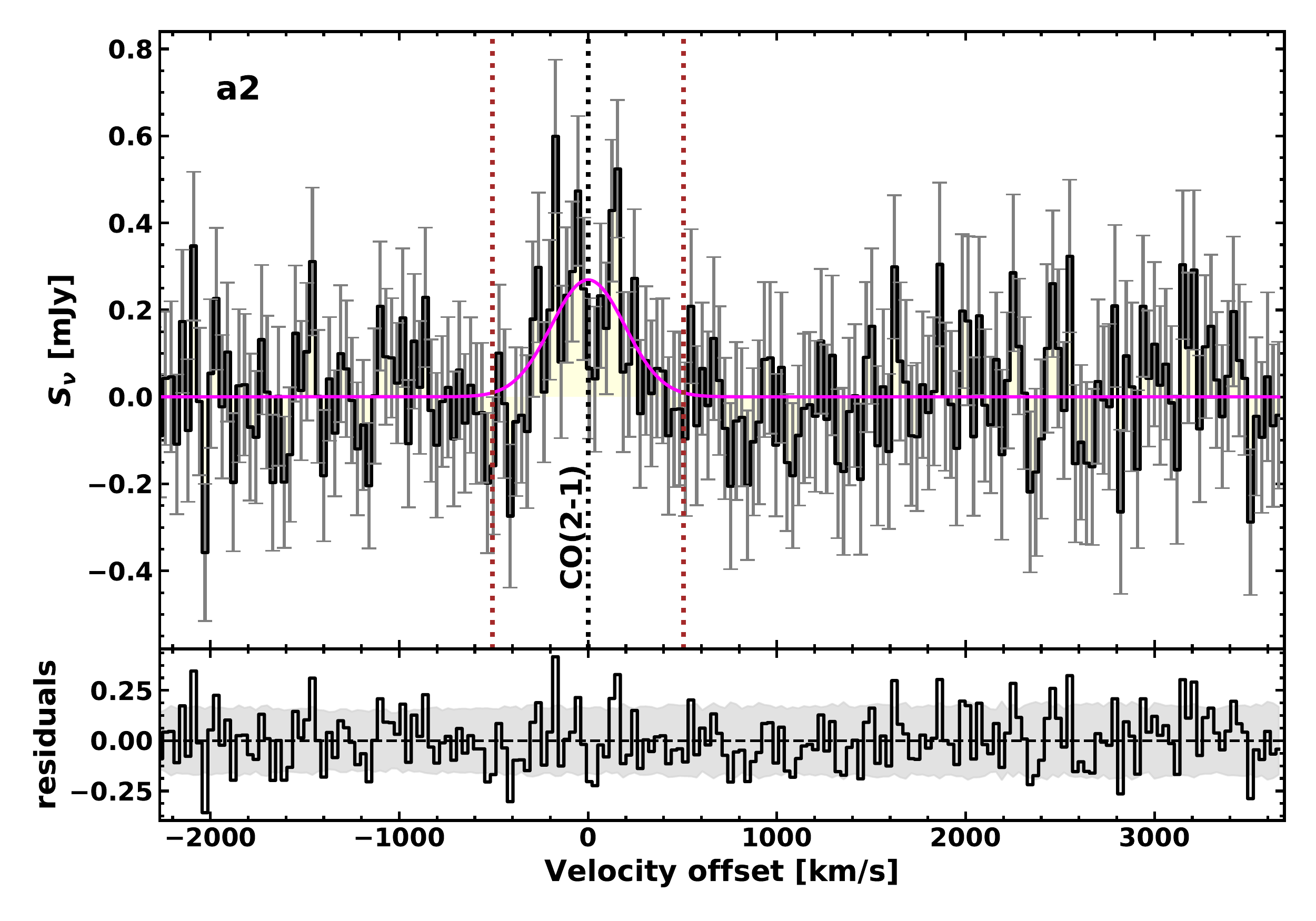}
\includegraphics[scale=0.3]{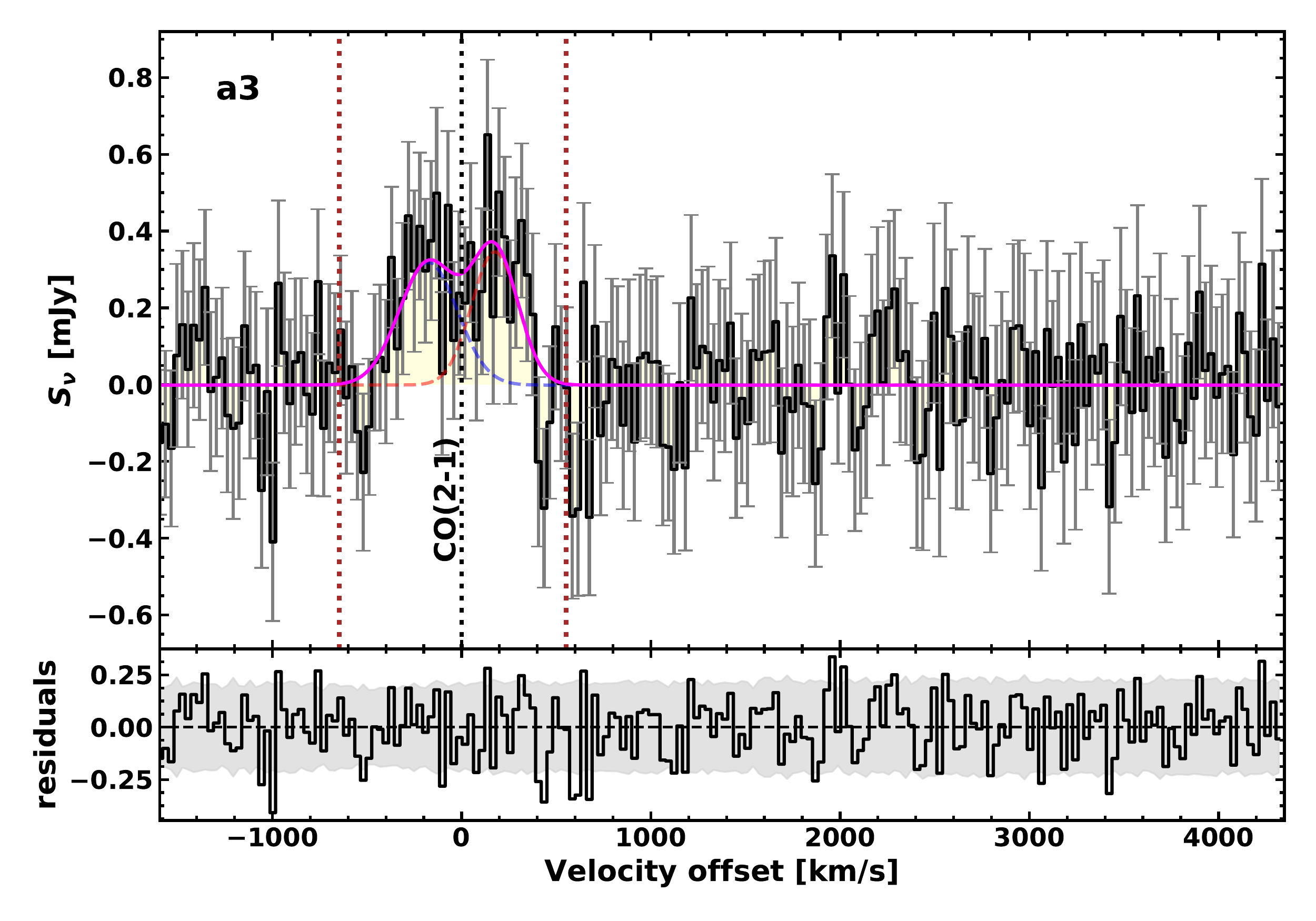}

        \caption{ALMA spectra of the four detected sources (channel width of 30 km/s). For each spectrum, the top panel shows the data extracted from the datacubes (solid black line) as a function of the radial velocity offset (increasing with redshift) from the detected line center. The gray bars represent the uncertainty per channel measured on the noise map. The solid magenta line marks the total model, i.e., the sum of the two components (dashed red and blue lines). The rest-frame zero-velocity is set to the central frequency of the fitted line (dotted black vertical line). For two-component fits, the central frequency of the line is set to the mean of the central frequencies of the two components. The vertical dotted brown lines delimit the velocity range within which we computed the moments. For the two-component fits, the range spans from the low-velocity side of the blueshifted to the high-velocity side of the redshifted component. The source ID is reported in the top left corner. Bottom panels: Residuals when the total model is subtracted from the data. The shaded gray region shows the rms per channel, indicating the 1$\sigma$ range.
        }
        
        \label{fig:spectra}
\end{figure*}

Using the S/N cube, we derived a 3$\sigma$ cube (i.e., the pixels of the original datacube were blanked except for those with S/N $\geq$ 3) that we used to define the region where the source spectra were extracted. Then, we performed a Markov chain Monte Carlo (MCMC) fit of the integrated spectra exploiting the {\sc{emcee}} Python package \citep{foreman_2013}, using either a single or double Gaussian model, depending on the observed line profile (see Appendix \ref{app:specandmom} for more details about the spectrum extraction and fitting). The spectra and their fits are shown in Fig. \ref{fig:spectra}.
In Table \ref{tab:results} we show the results of the spectral fitting. We list the CO(2$\rightarrow$1) redshift $\mathrm{z_{CO}}$, the full width at half maximum (\textit{FWHM}), and peak flux density $S_{CO(2-1)}$ of each line component (Cols. 2-6). We note that all new members are at a lower redshift than \object{$a0$}, and that the CO-derived redshift of the FRII core is consistent within 1$\sigma$ with that of LUCI ($\mathrm{z_{opt}}= 1.6987 \pm 0.0002$).\\
The spectral fitting results were exploited to constrain the velocity range (set to 1.5 $\times$ FWHM of the Gaussian model, corresponding to $\sim$3.5$\sigma$ from the peak) in the 3$\sigma$ cube, within which we derived the 3$\sigma$ -moments. 

\section{Results}
\label{sec:results}
We calculated the line-integrated flux $I_{CO(2-1)}$ as $\frac{1}{2}\sqrt{\frac{\pi}{\ln{2}}}$ $\times$ $(S_{CO(2-1)}\times FWHM)$ and
$\frac{1}{2}\sqrt{\frac{\pi}{\ln{2}}}$ $\times$ $(S_{CO(2-1),red}\times FWHM_{red}$ + $S_{CO(2-1),blue}\times FWHM_{blue})$ for the lines that were fit with one and two Gaussian components. The line luminosity $L'_{CO(2-1)}$ is then derived as $3.25 \times 10^{7}$ $\nu_{obs}^{-2}$ $(1+\mathrm{z_{CO}})^{-3}$ $D_L^2~ I_{CO}$  $\mathrm{K~km~s^{-1}~pc^2}$ , where $\nu_{obs}$ is the line observed frequency (in GHz) and $D_L$ is the luminosity distance (in Mpc; \citealt{solomon_1992}).
We derived the molecular hydrogen mass as $M_{H_2}$ = $(\alpha_{CO}/r_{21})$ $L'_{CO(2-1)}$ $M_{\odot}$, where $\alpha_{CO}$ is the assumed luminosity-mass conversion factor in units of $\mathrm{~M_{\odot}~(K~km~s^{-1}~pc^2)^{-1}}$ , and $r_{21}$ is the ratio between the luminosities of the CO(2$\rightarrow$1) and CO(1$\rightarrow$0) transitions. We assumed $r_{21}=0.76$ from the study of star-forming disk galaxies at z=1.5 by \cite{daddi_2015}, and a Galactic $\alpha_{\rm CO}=4.3$ $\mathrm{~M_{\odot}~(K~km~s^{-1}~pc^2)^{-1}}$ (\citeauthor{bolatto_2013} \citeyear{bolatto_2013}; see also \citeauthor{dannerbauer_2009} \citeyear{dannerbauer_2009}). However, we note that a lower $\alpha_{\rm CO}$ is possible for the FRII because the contribution of the high SFR (Sect. \ref{subsec:FRII_core}) and AGN to the excitation of the gas may be non-negligible \cite[e.g.,][]{papadopoulos_2012}. We report the derived fluxes, luminosities, and masses in Cols. 7-9 of Table \ref{tab:results}, respectively.\\
The FRII core is the only source that is marginally resolved in the sample. We derived its angular size performing a 2D Gaussian fit of the 3$\sigma$ moment 0 with the CASA task {\sc{imfit}}. We obtained a major axis $a =$ $3.2 \pm 2.0$ arcsec and a minor axis $b =$ $2.0 \pm 1.6$ arcsec (deconvolved from the beam), corresponding to $27 \pm 17$ kpc and $17 \pm 13$ kpc, respectively. Despite the large uncertainties, this suggests that the molecular gas extends across a significantly larger ($\sim3\times$) area than that of the FRII rest-frame optical emission, whose $H$-band derived major (minor) axis is ${\sim} 17$ (8.5) kpc. As a final remark, we also detect the ${\sim}$3 mm continuum emission from the nucleus and the western lobe of the FRII; these results will be presented in a future work focused on the multiband properties of the ALMA detected sources.

\begin{table*}
\caption{\label{tab:results} Summary of the fit results and molecular masses.
}
\centering
\resizebox{\hsize}{!}{
\begin{tabular}{cccccccccc}
\hline\hline

ID         &  $\mathrm{z_{CO}}$             & $FWHM_{red}$ & $FWHM_{blue}$ & $S_{CO(2-1),red}$ & $S_{CO(2-1),blue}$  &    $I_{CO(2-1)}$  & $L'_{CO(2-1)}$             & $M_{H_2}$   \\
           &                                 &        [km/s]         &        [km/s]           &       [mJy]                &          [mJy]               &        [Jy km/s]            & [$10^{9}~\mathrm{K ~km ~s^{-1} ~pc^2}$]   & [$10^{10}~(\frac{0.76}{r_{21}}\frac{\alpha_{CO}}{4.3})~ \mathrm{M_{\odot}}$]  \\  
(1)        &          (2)                    &      (3)              &      (4)        &        (5)                 &      (6)                     &       (7)                   &    (8)                                    &        (9)       \\      
\hline     
\object{$a0$}  & $1.6984^{+0.0003}_{-0.0004}$    &   $366^{+124}_{-89}$  &  $452^{+104}_{-90}$    &  $1.04^{+0.13}_{-0.11}$         & $1.07^{+0.11}_{-0.10}$       & $0.88^{+0.06}_{-0.06}$     & 33.6$\mathrm{^{+2.2}_{-2.3}}$      & 19.3$\mathrm{^{+0.1}_{-0.1}}$     \\                     
\object{$a1$}  & $1.6966^{+0.0001}_{-0.0001}$    &   \multicolumn{2}{c}{$133^{+43}_{-33}$}     &  \multicolumn{2}{c}{$0.51^{+0.14}_{-0.13}$}          &  $0.07^{+0.02}_{-0.01}$    & 2.6$\mathrm{^{+0.7}_{-0.6}}$         & 1.5$\mathrm{^{+0.4}_{-0.4}}$       \\                         
\object{$a2$}  & $1.6925^{+0.0004}_{-0.0005}$    &  \multicolumn{2}{c}{$447^{+97}_{-81}$}   &  \multicolumn{2}{c}{$0.28^{+0.05}_{-0.06}$}   &            $0.12^{+0.03}_{-0.03}$    & 4.7$\mathrm{^{+1.1}_{-1.1}}$        & 2.7$\mathrm{^{+0.6}_{-0.6}}$      \\                         
\object{$a3$}  & $1.6864^{+0.0006}_{-0.0006}$   & $304^{+237}_{-127}$ &  $365^{+201}_{-156}$     &  $0.33^{+0.10}_{-0.12}$  & $0.32^{+0.13}_{-0.12}$       &  $0.22^{+0.04}_{-0.03}$    & 8.4$\mathrm{^{+1.5}_{-1.3}}$           & 4.8$\mathrm{^{+0.8}_{-0.7}}$      \\                        
\hline
\end{tabular}
}
\tablefoot{
(1) ID of the source. (2) Redshift of the line, set equal to the mean of the blue- and redshifted components for the two-component fits. (3) and (4) \textit{FWHM} of the red- and blueshifted component, respectively. (5) and (6) Flux density peak of the red- and blueshifted components, respectively. (7) Integrated flux of the line. (8) Line luminosity. (9) Molecular gas mass. For \object{$a1$} and \object{$a2$} we performed a single-component fit.

}
\end{table*}

\section{Discussion and conclusions}
\label{sec:disc_conc}

\subsection{Gas-rich members and global properties of the system}

\begin{figure}[h!]
        \centering
\resizebox{\hsize}{!}
{\includegraphics[]{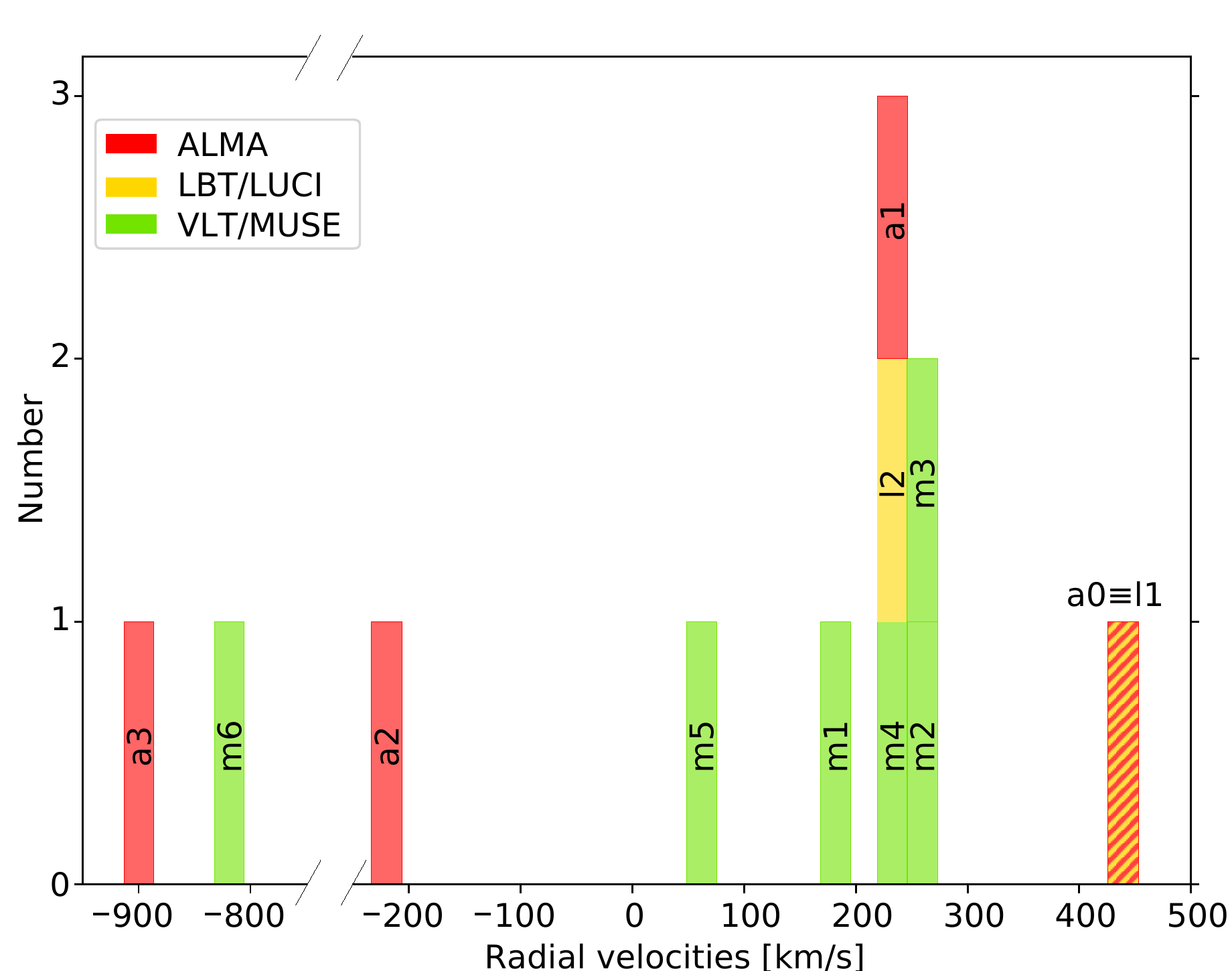}}

        \caption{Velocity distribution of all the known members of the system with respect to the mean redshift z=1.694, color-coded by the discovery instrument. The velocity bins are 30 km/s wide, equal to the mean uncertainty of the source redshift. \object{$a0$} and \object{$l1$} correspond to the same source, the FRII core.}
        
        \label{fig:vel_dist}
\end{figure}

We detected the FRII core (\object{$a0$}) and three new members of the large-scale structure (\object{$a1$}, \object{$a2$}, and \object{$a3$}). Three galaxies (\object{$a0$}, \object{$a1,$} and \object{$a2$}) lie within ${\sim}$80 kpc from each other, while \object{$a3$} is located at ${\sim}$650 kpc from the FRII core. 
Interestingly, \object{$a1$}, \object{$a2,$} and \object{$a3$} are all distributed westward of the FRII. The MUSE sources are all located eastward because the MUSE FoV covers only this part of the field (cyan box in Fig. \ref{fig:HST+srcs}, left). The spatial distribution of structure members around the FRII primarily suggests that the FRII may evolve into the brightest cluster galaxy (BCG) of a future cluster.
All the ALMA-detected sources show very red optical counterparts (Col. 5 of Table \ref{tab:sample}), indicating that they are dusty star-forming systems. Exploiting the Schmidt-Kennicutt \cite[SK,][]{kennicutt_1998} relation between the SFR surface density and the gas surface density $\Sigma_{SFR}$ = $2.5 \times 10^{-4}$ $\Sigma_{gas}^{1.4}$, we derived the SFRs for \object{$a1$}, \object{$a2,$} and \object{$a3$}. These SFRs vary from $\sim 5-50~M_\odot$/yr to $\sim 20-100~M_\odot$/yr, assuming for the size of the star-forming region as lower limit the optical size derived from the HST/F160w image (i.e., $\sim$ 46, 47, and 90 kpc$^2$ for \object{$a1$}, \object{$a2,$} and \object{$a3$}, respectively) and as upper limit the synthesized beam of the CO observations. It is interesting to note that \object{$a3$}, which features the highest gas mass and SFR after \object{$a0$}, is associated with a blend of two optical sources (HST image, Fig. \ref{fig:HST+srcs}) that may be ascribed to a major merger. In addition, we note that the bulk of the CO emission corresponds to the fainter optical source. Conservatively, in the optical size calculation, we considered  only the source corresponding to the bulk of the CO emission.\\
We did not detect any CO emission from the previously known MUSE sources. This is likely the consequence of selection effects because MUSE detects  unobscured sources and ALMA typically detects  obscured ones.
Based on the ALMA sensitivity limits at the source positions, and assuming a line width of 225 km/s (i.e., the mean of the fit components for \object{$a1$}, \object{$a2$} and \object{$a3$}), we derived 3$\sigma$ molecular mass upper limits for the MUSE sources, spanning $M_{H_2}$ $\leq$ $2.8-4.8\times$ $10^{10}~\mathrm{M_\odot}$, which do not rule out large amounts of molecular gas in the MUSE sources as well, however.\\
The mean redshift of the system (including all the known members) is z = 1.694 $\pm$ 0.001. The velocity distribution of the galaxies with respect to the mean redshift is shown in Fig. \ref{fig:vel_dist}. From the overall velocity distribution we derive a line-of-sight (LOS) velocity dispersion of ${\sim}$440 km/s. We note that the FRII, while showing the highest velocity, has a velocity offset of only ${\sim}$200 km/s from the distribution peak, in agreement with the hypothesis that it may become the dynamical center of the future cluster.

As for the system total mass, we first derived the virial mass from the LOS velocity dispersion following the procedure of \cite{cucciati_2018}, finding $M_{sys,vir} {\sim}6 \times 10^{13}~\mathrm{M_\odot}$. Considering that the system is probably far from being virialized, we provided a second estimate based on simple considerations on the overdensity. From the MUSE galaxy overdensity, located eastward of the FRII, \citetalias{gilli_2019} derived $M_{sys, MUSE} \gtrsim 1.5 \times 10^{13}~\mathrm{M_\odot}$. Based on the ALMA source positions, we can infer that the system extends across similar areas at either side of the FRII host. Under the minimal hypothesis that eastward and westward overdensities are the same, we can derive $M_{sys} \gtrsim 2 ~\times~ M_{sys, MUSE}$, that is, $M_{sys} \gtrsim 3 ~\times~ 10^{13}~\mathrm{M_\odot}$.
In addition, the photometric redshifts derived for a sample of X-ray obscured AGN distributed across the ${\sim}17'\times 17'$ \textit{Chandra} field suggest that the system extends on scales significantly larger than $\sim$1 Mpc \citep{peca_2020}.
We also note that the most massive dark matter halo is likely assembling around the FRII, that is, outside the MUSE FoV. Considering that the ratio between the stellar mass of the central galaxy in a halo and the halo mass is found to be at most ${\sim}2.5\%$ at z$\sim$2 \citep{behroozi_2013}, and that the FRII host stellar mass is $M_{\ast} {\sim} 3 \times 10^{11}~\mathrm{M_\odot}$, we derive a halo mass of $\gtrsim 1.2 \times 10^{13}~\mathrm{M_\odot}$. On the basis of cosmological simulations, \cite{chiang_2013} have studied the expected evolution of the most massive halo of a cluster progenitor as a function of redshift. The mass found for the FRII halo places our system among those objects that would evolve into a z = 0 cluster of at least ${\sim}1.4-3 \times 10^{14}~\mathrm{M_\odot}$. Furthermore, \cite{chiang_2013} presented a correlation between the galaxy overdensity profile and the future cluster mass (see their Fig. 10). By assuming again that the ALMA and MUSE galaxy overdensities are the same (i.e., $\delta_{gal}\sim 7.4$; \citetalias{gilli_2019}), we found that in a volume similar to that observed by ALMA ($\sim$15 Mpc$^3$), our system at z$\sim$2 will likely evolve into a cluster of ${\sim}1 \times 10^{15}~\mathrm{M_\odot}$ at z = 0.

\subsection{FRII host galaxy}
\label{subsec:FRII_core}
\begin{figure}[h!]
        \centering
{\includegraphics[scale=0.5]{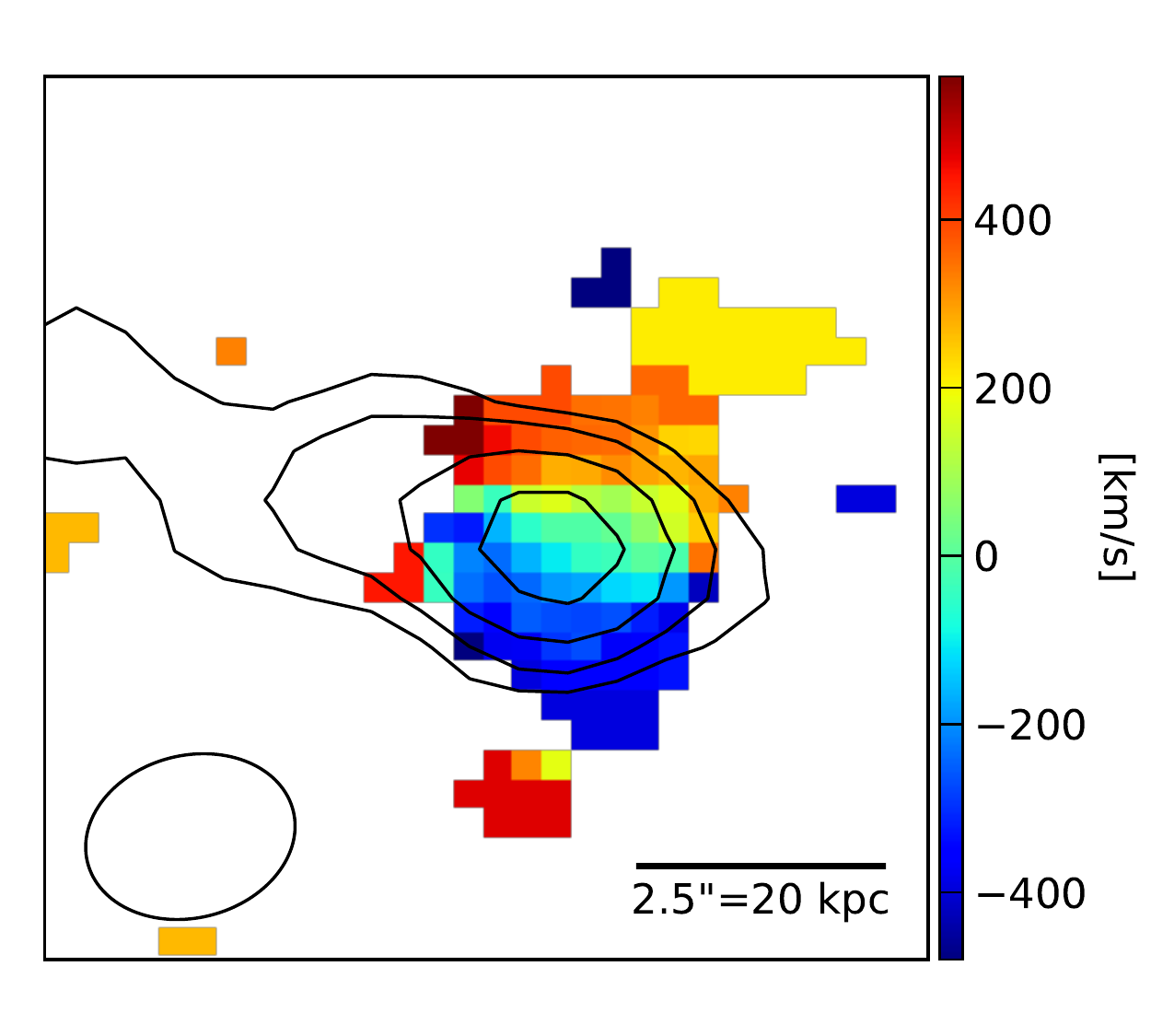}}

        \caption{FRII host galaxy (\object{$a0$}) line velocity map (moment 1, with respect to the line central frequency of the fit marked by the vertical dotted black line in Fig. \ref{fig:spectra}). North is up, and east is to the left. The open black ellipse is the restoring beam. The VLA contours as in Fig. \ref{fig:HST+srcs} are reported in black.}
        
        \label{fig:a0_moment1}
\end{figure}

The FRII core is by far the strongest CO emitter, unveiling a large molecular gas reservoir (${\sim}2\times 10^{11}~\mathrm{M_\odot}$) that surrounds the radio galaxy host. This gas component is one of the most massive reported so far for an HzRG nucleus in an overdense region. Other examples of large gas reservoirs detected around similar objects are the Spiderweb (${\sim}2 \times 10^{10}~(\alpha_{CO}/0.8)~\mathrm{M_\odot}$ at z=2.2; \citeauthor{emonts_2018} \citeyear{emonts_2018}), SpARCS J1049+56 ($M_{H2}$ $\sim$ $1\times 10^{11}~(\alpha_{CO}/0.8)~\mathrm{M_\odot}$ at z=1.7; \citeauthor{webb_2017} \citeyear{webb_2017}; see also \citeauthor{castignani_2020} \citeyear{castignani_2020}), and Candels-5001 ($M_{H2}$ $\sim$ $2-6 \times 10^{11}~(\alpha_{CO}/10)~\mathrm{M_\odot}$ at z=3.47; \citeauthor{ginolfi_2017} \citeyear{ginolfi_2017}) BCGs.\\
We again used the SK relation to derive an SFR of ${\sim}$ 200 -- 600 $M_\odot$/yr, assuming for the star-forming region a size ranging from that of the optical rest-frame emission (derived from the HST/F160w image, ${\sim}110~\mathrm{kpc^2}$) to that of the molecular gas ($\pi \times a/2 \times b/2$, ${\sim}360~\mathrm{kpc^2}$). Coupled with the high radio-to-X-ray luminosity, stellar mass, central position in the structure, and gas reservoir of the source (significantly higher than those of the other members), this strongly favors a scenario in which the FRII is the BCG progenitor of the cluster, caught during an active phase of super massive black hole and host galaxy growth, likely induced by the extremely gas-rich environment.\\
From the size of the source and its mass, we attempted an estimate of the interstellar medium (ISM) column density ($N_{H,ISM}$) along the LOS, and compared it with that derived from the X-ray spectrum ($N_{H,X}{\sim}~1.5 \times 10^{24}~ \mathrm{cm^{-2}}$; \citetalias{gilli_2019}) to evaluate the contribution of the host galaxy to the nuclear obscuration. Because of the large uncertainty on the source size, the inclination angle and disk thickness 
of the molecular gas are unconstrained (i.e., the relative errors are $\geq$1).
We therefore decided to derive an upper limit for the column density assuming an edge-on disk (i.e., the inclination angle between the rotation axis and the LOS is $\theta = 90^\circ$). As for the geometry, we assumed a thin disk having the diameter equal to $a$ and constant height equal to 1 kpc, similar to the typical scale height of high-z spirals and chain galaxies \citep{elmegreen_2006}. We found $N_{H,ISM}{\sim}~5.5 \times 10^{23} ~\mathrm{cm^{-2}}$. This column density is considerable, suggesting that the ISM may contribute significantly to the total nuclear obscuration measured in the X-rays, in addition to a small, parsec-scale absorber, possibly arranged in a torus-like geometry around the nucleus \cite[as postulated in the Unified Schemes;][]{urry_1995}. 
From the flux ratio of the two radio jets, \citetalias{gilli_2019} derived the inclination angle between the approaching radio jet and the LOS, finding that the system is seen almost in the plane of the sky (i.e, $\theta{\sim}70^\circ-80^\circ$). Under the assumption that the radio jet and the torus are coaxial, the high $N_{H,X}$ value measured by \citetalias{gilli_2019} is consistent with an edge-on orientation of the torus. From the velocity map in Fig. \ref{fig:a0_moment1} we observe a velocity gradient that points toward a structure that rotates perpendicularly to the radio jet (shown by the 1.4 GHz VLA black contours), at least in projection. This seems to indicate that the inner accretion disk (and torus) and the galaxy-scale disk have a common rotation axis. However, existing 3D modeling of local radio galaxies (\citealt{verdoes-kleijn_2005}; Ruffa et al. in prep.) show that the relative inclination angle between the jet and the (sub-)kpc molecular (and/or dust) disks can vary over a wide range of values, and that a fully axis-symmetric scenario may be too simplistic. In the moment 1 map we also note two peculiar kinematical features (northwest and southeast of the nucleus), which appear inconsistent with rotation, and possibly associated with noncircular motions. Further observations at higher resolution and S/N are needed to unveil the nature of these features.

\begin{acknowledgements}
     QD acknowledges F. Loiacono, I. Ruffa and R. Paladino for their help. IP acknowledges support from INAF under the SKA/CTA PRIN ``FORECaST'' and the PRIN MAIN STREAM ``SauROS'' projects. PT acknowledges support from the INAF PRIN-SKA 2017 program 1.05.01.88.04 (ESKAPE). This letter makes use of the following ALMA data: {2018.} {1.}{01601.}S. ALMA is a partnership of ESO (representing its member states), NSF (USA) and NINS (Japan), together with NRC (Canada), MOST and ASIAA (Taiwan), and KASI (Republic of Korea), in cooperation with the Republic of Chile. The Joint ALMA Observatory is operated by ESO, AUI/NRAO and NAOJ.
\end{acknowledgements}

\bibliographystyle{aa} 
\bibliography{J1030_ALMA} 

\appendix

\section{Additional material}

\subsection{S/N cube and source detection code}
\label{app:det_code}
In order to produce the S/N cube, we first generated a noise cube in which each pixel value is the rms of the pixels in a region centered in it. As for the size of this region, we found that a 10-beam area is small enough to trace the local variations of the noise and is large enough to have sufficient statistics. For each pixel the rms is calculated to convergence: the process iteratively excludes the pixels above 3 $\times$ rms in the surrounding 10-beam region in order to remove the effect of real sources in the rms evaluation. Then, the S/N cube is produced from the ratio between the datacube and the noise cube.\\
The detection code works as follows: first, it scans each pixel of each channel (i.e., each spaxel) of the S/N cube and searches for a given number $\mathrm{N_{ch}}$ of contiguous channels (named ``sliding window'') above a given S/N threshold ($\mathrm{S/N_{w}}$) and selects them. Then the code sums the S/N of the $\mathrm{N_{ch}}$ channels in each selected sliding window. A candidate detection is then associated with the pixel with the
highest S/N in the sliding window with the highest sum of S/N over its $\mathrm{N_{ch}}$ channels, only if the pixel has S/N $>$ 3. Second, all candidate detections with a number of spatially contiguous detected pixels lower than a given number $\mathrm{N_{px}}$ are rejected. Finally,  the reliability of the detected sources (``positive detections'') is assessed by running the code on the negative S/N cube (i.e. a cube in which each pixel has the same value, but inverted sign). These ``negative detections'' provide us with information about the incidence and S/N distribution of spurious detections. A conservative detection criterion is to select only the sources found with any given parameter set for which no negative detection are found. We performed the source research for all possible combinations of $\mathrm{S/N_{w}} = 1.5, 2.0, 2.5, 3.0$, $\mathrm{N_{ch}} = 2, 3, 4$ (corresponding to 60, 90, and 120 km/s, respectively) and $\mathrm{N_{px}} = 2, 3, 4$. The maximum values of $\mathrm{N_{ch}}$ and $\mathrm{N_{px}}$ (i.e., the most stringent adopted criteria) are such that the code begins to recover only the brightest sources, and no additional detections are found. The searching area was cropped to the \textit{HPBW} of the ALMA FoV in order to exclude the field outskirts where the noise increases very steeply and possibly produces artifacts. From all the candidate detections that we obtained with a given combination of $\mathrm{S/N_{w}}$, $\mathrm{N_{ch}}$ , and $\mathrm{N_{px}}$, we then excluded those for which we obtained at least one negative detection for the same parameter combination. We found that the code is highly reliable (i.e., no negative detections found) down to $\mathrm{S/N_{w}} = 2.0$ for $\mathrm{N_{ch}}$ = 4 (120 km/s) and down to $\mathrm{S/N_{w}} = 2.5$ for $\mathrm{N_{ch}}$ = 3 (90 km/s), requiring at least $\mathrm{N_{px}}$ = 3. After the reliability step, the only remaining detections for all the possible parameter combinations are \object{$a0$}, \object{$a1$}, \object{$a2,$} and \object{$a3$}; in this letter we report only these secure four detections that satisfy the adopted stringent reliability criterion.

\subsection{Spectrum extraction and fitting}
\label{app:specandmom}
In order to extract the spectra, we created a blanked 3$\sigma$ cube from the original datacube, masking all pixels with S/N $<$ 3. We first computed the 3$\sigma$ integrated flux map (moment 0) and velocity map (moment 1) for each source over the whole frequency range covered by the cube. Then, we extracted the integrated spectra of each source from the original datacube in a region drawn considering all the contiguous pixels around the peak position in the 3$\sigma$ moment 0. In this process, we exploited the moment 1 maps to reject the pixels whose velocities were largely inconsistent  with the observed line velocity range (i.e., displaced by $>$ 1000 km/s). For the only resolved source (\object{$a0$}) we measured the flux density per channel (i.e., the spatially integrated surface brightness); for the remaining sources, all unresolved, we took the mean per channel. The error per channel was measured by taking the mean of the rms over the same region of the noise cube.\\
As for the line MCMC fitting, a model with two Gaussian components was used for \object{$a0$} and  \object{$a3$}, which clearly show a double peaked feature in the spectrum, and (for \object{$a0$}) a velocity gradient in the moment 1 map is present. As for and \object{$a1$} and \object{$a3$}, only one Gaussian component was used in the fit.

\end{document}